\documentstyle[12pt]{article}
\def\be{\begin{eqnarray}}
\def\ee{\end{eqnarray}}
\def\ba{\begin{array}}
\def\ea{\end{array}}
\def\G{{\cal G}}
\def\B{{\cal B}}
\def\A{{\cal A}}
\def\M{{\cal M}}
\def\T{{\cal T}}
\def\L{{\cal L}}
\def\X{{\cal X}}
\def\Z{{\cal Z}}
\def\D{{\cal D}}
\def\H{{\cal H}}
\def\C{{\cal C}}
\def\Q{{\cal Q}}
\def\K{{\cal K}}
\def\N{{\cal N}}
\begin{document}
\begin{center}
{\bf\LARGE {3D heterotic string theory:\\
         \vskip 3mm
	    new approach  and extremal solutions}}
\end{center}
\vskip 25mm
\begin{center}
{\bf \large {Oleg V. Kechkin}}
\end{center}
\begin{center}
DEPNI, Institute of Nuclear Physics\\
M.V. Lomonosov Moscow State University\\
119899 Moscow, Vorob'yovy Gory, Russia\\
e-mail:\, kechkin@depni.npi.msu.su
\end{center}
\vskip 25mm
\begin{abstract}
We develop a new formalism for the bosonic sector of low-energy heterotic
string theory toroidally compactified to three dimensions. This formalism is
based on the use of some single non-quadratic real matrix potential which
transforms linearly under the action of subgroup of the three-dimensional
charging symmetries. We formulate a new charging symmetry invariant approach for
the symmetry generation and straightforward construction of asymptotically flat
solutions. Finally, using the developed approach and the established formal
analogy between the heterotic and Einstein-Maxwell theories, we construct a
general class of the heterotic string theory extremal solutions of the
Israel-Wilson-Perjes type. This class is asymptotically flat and charging
symmetry complete; it includes the extremal solutions constructed before
and possesses the non-trivial bosonic string theory limit.
\end{abstract}
\vskip 10mm
\begin{center}
\end{center}
\renewcommand{\theequation}{\thesection.\arabic{equation}}
\newpage
\section{Introduction}
\setcounter{equation}{0}

Field theory limits of superstring theory include some concrete modifications of
the classical General Relativity \cite{Kir}. These effective field
theories describe a dynamics of some set of the superstring excitation modes
restricted by the corresponding limit conditions. In the low-energy limit of
heterotic string theory one deals with the zero-mass modes, which bosonic sector
includes the dilaton, Kalb-Ramond, Abelian gauge and metric fields. These fields
live in the multidimensional space-time and interact in the supergravity
controlled form (see \cite{GSW} and references therein).

A solution space of the low-energy limit of heterotic string theory had been
extensively studied during last several years \cite{SenSol1}, \cite{SenSol2},
\cite{Cve1}, \cite{Cve2},
\cite{KalIWP}, \cite{Youm}, \cite{OurSol}; its investigation actively continues
at the present time. The main part of the both
physically and mathematically interesting solutions had been found for the
theory toroidally compactified to three and lower dimensions (see, however, some
higher-dimensional examples in \cite{Mal} and works referred there). The reason
for such activity generation is closely related to the fact of hidden symmetry
enhancement in the process of the toroidal compactification \cite{MahSch},
\cite{Sen4}. The qualitative important new situation takes place in the case of
toroidal compactification to three dimensions, when the heterotic
string theory becomes the three-dimensional symmetric space model coupled to
gravity \cite{Sen3}. Such a theory belongs to a class described in \cite{BGM} and
possesses the complete integrability property after the following reduction to
two dimensions \cite{Sch}, \cite{BKR}. In this sense the heterotic string theory
is also of the same type as the Einstein, Kaluza-Klein and Einstein-Maxwell
theories \cite{BZ}, \cite{Kin}, \cite{Al}, \cite{EGK}.

In our approach we consider the three-dimensional heterotic string theory in
a remarkable explicit Einstein-Maxwell form. In \cite{HKMEP} it was
established a new form of the null-curvature matrix representation of the
theory. This form is based on the use of the Ernst matrix potentials \cite{HKCS}.
The Ernst matrix potential formulation is the straightforward matrix
generalization of the conventional one of the stationary Einstein-Maxwell theory
\cite{ErnstEM} to the heterotic string theory case. In fact this generalization
includes the Einstein-Maxwell theory as some special case (see \cite{HKCS} and
this work below), and gives a convenient and simple toll for the symmetry analysis
of the theory. In \cite{HKCS} we have classified the three-dimensional group of
hidden symmetries in the matrix-valued Einstein-Maxwell form; namely we have
separated it to the shift, electric-magnetic rotation, Ehlers and Harrison type
parts of transformations (compare with the ones in \cite{Kin}, \cite{Ehl}; see
also \cite{EMT}). Also it was constructed the full subgroup of charging
symmetries, i.e. the total set of the symmetry transformations which preserve the
asymptotic flatness property of the solutions (see in \cite{EMT} references for
the Einstein-Maxwell theory analogies). It was established a representation of the
theory which is linear in respect to the action of whole set of the charging
symmetry transformations.

In this article we continue and conclude this line of investigations of the
three-dimensional heterotic string theory. We develop in details the above
mentioned representation and formulate a new approach to the theory which is
based on the use of new single non-quadratic matrix potential. This potential
has the lowest possible matrix dimensionality compatible with the structure of
the symmetric space model of the theory and gives the powerful and the most
compact
tool for the theory investigation. We introduce three doublets of matrix
potentials defined in terms of this underlying above mentioned one. Any doublet
consists of one scalar and one vector matrix potential; both these potentials
undergo linear transformations when the charging symmetry subgroup acts. We
formulate a new method for the symmetry generation and straightforward
construction of
the solutions which guarantees the charging symmetry completeness property of
the result. Our approach is especially useful for the work with asymptotically
flat solutions, which attract the main interest in the physical applications of
the string theory (see \cite{Youm} for the string theory based black hole
physics). Also we establish a relation between our new representation with the
null-curvature matrix one. This gives some new simplifications and promising
possibilities in construction of the two-dimensional solutions using the inverse
scattering transform method. We hope to illustrate this statement in the nearest
future. In this article we apply the new formalism for straightforward
construction
of the charging symmetry invariant class of the asymptotically flat extremal
solutions of the Israel-Wilson-Perjes type. We `rediscover' the well-known
corresponding Einstein-Maxwell theory solution class \cite{IWP}, and give its
straightforward generalization to the heterotic string theory case. Our class
of solutions closely relates to the known in the literature \cite{KalIWP},
\cite{Sab}, \cite{OurIWP}, and also includes the non-trivial
subclass of the extremal bosonic string theory solutions, i.e. the solutions
without multidimensional Abelian gauge fields.

\section{3D heterotic string theory}
\setcounter{equation}{0}

The action for the bosonic sector of low-energy heterotic string theory reads
\cite{Kir}:
\be\label{h1}
S_D=\int d^DX |{\rm det}G_{MN}|^{\frac{1}{2}}\,e^{-\Phi}
\left ( R_D+\Phi_{,M}\Phi^{,M}-\frac{1}{12}H_{MNK}H^{MNK}-
\frac{1}{4}F^I_{MN}F^{I\,MN}\right ),
\ee
where $H_{MNK}=\partial_MB_{NK}-\frac{1}{2}A^I_MF^I_{NK}+\,{\rm cyclic}\,
\{M,N,K\}$ and $F^I_{MN}=\partial_MA^I_N-\partial_NA^I_M$. Here $X^M$ is the
M-th $(M=1,...,D)$ coordinate of the physical space-time of the signature
$(-,+,...,+)$, $G_{MN}$ is the metric, whereas $\Phi$, $B_{MN}$ and $A^I_M$ \,
$(I=1,...,n)$ are the dilaton, Kalb-Ramond and Abelian gauge fields.

As it was declared in Introduction, in this paper we consider the theory
(\ref{h1}) toroidally compactified to three dimensions. Let us briefly describe
this compactification (\cite{MahSch}, \cite{Sen4}, \cite{Sen3}). First of all,
let us put $D=d+3$ and denote $Y^M=X^m$\, $(m=1,...,d)$, and $x^{\mu}=X^{d+\mu}$
\, $\mu=1,2,3$. Second, let us separate all the field components to the scalar,
vector and $2$-rank tensor quantities in respect to transformations of the
coordinates $x^{\mu}$. As the result one has three scalar matrices $G$, $B$
and $A$ of the dimensions $d\times d$, $d\times d$ and $d\times n$ constructed
from the components $G_{mk}$, $B_{nk}$ and $A_{mI}=A^I_m$ respectively, and also
the scalar function
\be\label{h2}
\phi=\Phi-{\rm ln }|{\rm det}\,G|^{\frac{1}{2}}.
\ee
Then, there are three vector matrix columns $\vec V_1$, $\vec V_2$ and $\vec V_3$
of the dimensions $d\times 1$, $d\times 1$ and $n\times 1$. They read:
\be\label{h3}
&&V_{1\,m\mu}=G^{-1}_{mk}G_{k\,d+\mu},
\nonumber\\
&&V_{2\,m\mu}=B_{m\,d+\mu}-B_{mk}V_{1\,k\mu}+\frac{1}{2}A^I_mV_{3\,I\mu},
\nonumber\\
&&V_{3\,I\mu}=-A^I_{d+\mu}+A^I_mV_{1\,m\mu}.
\ee
Finally, there are two tensor fields, they consist of the non-matrix quantities
\be \label{h4}
&&h_{\mu\nu}=e^{-2\phi}\left [ G_{d+\mu\,d+\nu}-G_{mk}
V_{1\,m\mu} V_{1\,k\nu}\right ],
\nonumber\\
&&b_{\mu\nu}=B_{d+\mu\,d+\nu}-B_{mk} V_{1\,m\mu} V_{1\,k\nu}-
\frac{1}{2}\left [ V_{1\,m\mu} V_{2\,m\nu}-
V_{1\,m\nu} V_{2\,m\mu}\right ].
\ee
Let us now perform the toroidal compactification of the first $d$ dimensions. In
fact this procedure is equivalent to consideration of the special situation when
all the field components are the $Y^m$-independent. Thus, below all the quantities
are considered as the functions of the coordinates $x^{\mu}$. The use of the
motion
equations allows one to introduce the pseudoscalar fields $u, v$ and $s$ (they
are the $d\times 1$, $d\times 1$ and $n\times 1$ columns) accordingly the
relations:
\be \label{h5}
\nabla\times\vec V_1&=&e^{2\phi}G^{-1}
\left [\nabla u+(B+\frac{1}{2}AA^T)\nabla v+A\nabla s\right ],
\nonumber                            \\
\nabla\times\vec V_2&=&e^{2\phi}G\nabla v-
(B+\frac{1}{2}AA^T)\nabla\times\vec V_1+
A\nabla\times\vec V_3.
\nonumber \\
\nabla\times\vec V_3&=&e^{2\phi}
(\nabla s+A^T\nabla v)+A^T\nabla\times\vec V_1,
\ee
where all the vector operations are defined respectively the three-metric
$h_{\mu\nu}$. Then, the tensor field $b_{\mu\nu}$ is non-dynamical; following
\cite{Sen3} we put $b_{\mu\nu}=0$ in our analysis. This restriction does not
possess any new ones on the remaining dynamical quantities. The resulting
effective three-dimensional system coincides with some symmetric space model
coupled to gravity \cite{Sen3}. To express it in terms convenient for our
consideration, let us introduce the following matrices $\G$, $\B$ and $\A$
\cite{HKMEP}:
\be \label{h6}
\G=
\left(
\ba{cc}
-e^{-2\phi}+v^TGv&v^TG\cr
Gv&G
\ea
\right),
\quad
\B=
\left(
\ba{cc}
0&-w^T\cr
w&B
\ea
\right),
\quad
\A=\left(
\ba{c}
s^T+v^TA \cr
A
\ea
\right),
\ee
where $w=u+Bv+\frac{1}{2}As$. Then, let us define the $[2(d+1)+n]\times
[2(d+1)+n]$ matrix $\M$ as
\be \label{h7}
\M=
\left(
\ba{ccc}
\G^{-1}&\G^{-1}(\B+\T)&\G^{-1}\A\cr
(-\B+\T)\G^{-1}&(\G-\B+\T)\G^{-1}(\G+\B+\T)&(\G-\B+\T)\G^{-1}\A\cr
\A^T\G^{-1}&\A^T\G^{-1}(\G+\B+\T)&1+\A^T\G^{-1}\A
\ea
\right),
\ee
where $\T=\frac{1}{2}\A\A^T$. This matrix satisfies the restrictions
\be \label{h8}
\M^T=\M, \qquad \M\L\M=\M,
\ee
where
\be \label{h9}
\L=
\left (
\ba{ccc}
0&1&0\cr
1&0&0\cr
0&0&-1
\ea
\right),
\ee
so it parameterizes the symmetric space $O(d+1,d+1+n)/O(d+1)\times O(d+1+n)$.
The resulting three-dimensional theory can be expressed in terms of the
three-metric $h_{\mu\nu}$ and the matrix $\M$; the corresponding action reads:
\be \label{h10}
S_3=\int d^3x h^{\frac{1}{2}}\left ( -R_3+L_3\right ),
\ee
where
\be \label{h11}
L_3=\frac{1}{8}{\rm Tr}\left (\nabla\M\,\M^{-1}\right )^2.
\ee
At the first time the representation (\ref{h10})-(\ref{h11}) was considered in
\cite{Sen3} with another form of the null-curvature matrix $\M$. In \cite{GKMEP}
it was found a symplectic matrix representation for the special symmetric space
model with $d=n=1$. In \cite{OKU} one can find the unitary null-curvature matrix
for the system with $d=1, n=2$.

Now let us use the motion equations and Eq. (\ref{h8}) and introduce on shell
the vector matrix $\vec\Omega$ accordingly the relation
\be \label{h12}
\nabla\times\vec\Omega=\nabla\M\,\L\,\M;
\ee
then from Eq. (\ref{h8}) it follows that $\vec\Omega^T=-\vec\Omega$. Then, from
Eqs. (\ref{h6}), (\ref{h7}) and (\ref{h12}) one concludes, that the matrices
$\M$ and $\vec\Omega$ have the equivalent natural block matrix structure: they
are $5\times 5$ block matrices. Using the straightforward calculations one
can check that
\be \label{h13}
\vec V_1=\vec\Omega_{12}^T,\quad \vec V_2=\vec\Omega_{14}^T,\quad
\vec V_3=\vec\Omega_{15}^T,
\ee
where the indexes enumerate the corresponding matrix blocks. Let us also define
the following set of the scalar quantities:
\be \label{h14}
&&S_0=-\M_{11}, \quad S_1=\M_{22}-\M_{11}^{-1}\M_{12}^T\M_{12},
\nonumber\\
&&S_2=\M_{24}-\M_{11}^{-1}\M_{12}^T\M_{14},\quad
S_3=\M_{25}-\M_{11}^{-1}\M_{12}^T\M_{15}.
\ee
We state that these scalar and vector potentials can be effectively explored for
the construction of the all multidimensional fields of heterotic string theory
(\ref{h1}). Namely, for the $D=(d+3)$-dimensional line element one has:
\be\label{h15}
ds^2_{d+3}=(dY+V_{1\mu}dx^{\mu})^TS_1^{-1}(dY+V_{1\nu}dx^{\nu})+S_0ds_3^2,
\ee
where $Y$ is the $d$-dimensional coordinate column with the components $Y^m$ and
$ds_3^2=h_{\mu\nu}dx^{\mu}dx^{\nu}$. Then, for the matter fields one has the
following expressions:
\be\label{h16}
&&e^{\Phi}=\left | S_0\,{\rm det}\,S_1\right |^{\frac{1}{2}},
\nonumber\\
&&B_{mk}=\frac{1}{2}\left ( S_1^{-1}S_2-S_2^TS_1^{-1}\right )_{mk},
\nonumber\\
&&B_{m\,d+\nu}=\left \{ V_{2\nu}+\frac{1}{2}\left ( S_1^{-1}S_2-
S_2^TS_1^{-1}\right )V_{1\nu}-S_1^{-1}S_3V_{3\nu}\right \}_{m},
\nonumber\\
&&B_{d+\mu\,d+\nu}=\frac{1}{2}\left [ V_{1\mu}^T\left ( S_1^{-1}S_2-
S_2^TS_1^{-1}\right )V_{1\nu}+V_{1\mu}^TV_{2\nu}-V_{1\nu}^TV_{2\mu}\right ],
\nonumber\\
&&A^I_m=\left ( S_1^{-1}S_3\right )_{mI},
\nonumber\\
&&A^I_{d+\mu}=\left ( -V_{3\mu}+S^T_3S^{-1}_1V_{1\mu}\right )_I.
\ee
Eqs. (\ref{h13})-(\ref{h16}) allow one to translate any solution of the
three-dimensional problem (\ref{h10}), (\ref{h11}) to the form of physical
fields of the heterotic string theory (\ref{h1}). They are especially useful in
framework of the solution construction approaches based on the use of the
null-curvature matrix $\M$. This situation takes place when one uses the
Kramer-Neugebauer geodesic method \cite{EMT} or the Belinsky-Zakharov inverse
scattering transform technique \cite{BZ}. Eqs. (\ref{h13})-(\ref{h16}) also play
an important role in the our new approach to the three-dimensional heterotic
string theory which is developed in the next section.

\section{New approach}
\setcounter{equation}{0}
For the following analysis it is necessary to introduce a pair of the Ernst matrix
potentials \cite{HKCS}. These are the matrices $\X$ and $\A$,
where
\be\label{n1}
\X=\G+\B+\frac{1}{2}\A\A^T,
\ee
and $\A$ is the potential introduced in the previous section. In terms of
these potentials the three-dimensional Lagrangian $L_3$ takes the following form:
\be\label{n2}
L_3={\rm Tr}\,\left [
\frac{1}{4}\left(\nabla\X-\nabla\A\A^T
\right)\G^{-1}
\left(\nabla \X^T-\A\nabla \A^T\right)
\G^{-1}+\frac{1}{2}\G^{-1}\nabla\A\nabla\A^T
\right ];
\ee
here $\G=\frac{1}{2}(\G+\G^T-\A\A^T)$. The simplest solution of this
three-dimensional theory, which corresponds to the empty Minkowskian space-time,
is given by the matrices $\X_0=\Sigma={\rm diag}\,(-1,-1;1,...,1)$, $\A_0=0$
and by the three-dimensional metric $h_{\mu\nu}=\delta_{\mu\nu}$, see Eqs.
(\ref{h6}), (\ref{n1}). Following the conventional terminology of General
Relativity (\cite{EMT}) we name solutions `asymptotically flat' if
$\X\rightarrow \Sigma$, $\A\rightarrow 0$ when the three-point with the
coordinates $x^{\mu}$ tends to the spatial infinity (note, that time is taken as
one of the compactified dimensions). Let us now introduce the following pair of
matrix potentials:
\be\label{n3}
\Z_1=2\left ( \X+\Sigma\right )^{-1}-\Sigma, \quad
\Z_2=\sqrt 2\left ( \X+\Sigma\right )^{-1}\A.
\ee
The map $(\X,\A)\rightarrow (\Z_1,\Z_2)$ coincides with the inverse one; the
similar substitution is familiar in the stationary Einstein-Maxwell theory
\cite{Maz}. Our new approach is based on the use of the single $(d+1)\times
(d+1+n)$ matrix potential $\Z$, where
\be\label{n4}
\Z=(\Z_1\,\,\Z_2).
\ee
This matrix potential is quadratic for the bosonic string and non-quadratic for
the heterotic string theories. Using tedious but straightforward
calculations and Eqs. (\ref{n2})-(\ref{n4}), one can prove that in terms of the
potential $\Z$ the three-dimensional matter Lagrangian reads:
\be\label{n5}
L_3={\rm Tr}\,\left [\nabla\Z\left (\Xi-\Z^T\Sigma\Z\right )^{-1}
\nabla\Z^T\left (\Sigma-\Z\Xi\Z^T\right )^{-1}\right ],
\ee
where $\Xi={\rm diag}\,(-1,-1;1,...,1)$ is the $(d+1+n)\times (d+1+n)$ matrix.
Eqs. (\ref{h10}) and (\ref{n5}) define the $\Z$-based formalism completely. The
corresponding motion equations are:
\be\label{n6}
&&\nabla^2\Z+2\nabla\Z\Xi\Z^T\left (\Sigma-\Z\Xi\Z^T\right )^{-1}\nabla\Z=0,
\nonumber\\
&&R_{3\,\mu\nu}={\rm Tr}\,\left [\Z_{,(\mu}
\left (\Xi-\Z^T\Sigma\Z\right )^{-1}\Z^T_{,\nu)}
\left (\Sigma-\Z\Xi\Z^T\right )^{-1}\right ].
\ee
In our plans is to develop the $\Z$-formalism in details; namely, we would like
to obtain its explicit relation to the null-curvature matrix representation, the
$\Z$-based scheme of calculation of the multidimensional field components and
also the representation of the all hidden symmetries in terms of $\Z$. To realize
this program, let us express the matrix $\M$ from Eq. (\ref{h7}) in terms of the
Ernst matrix potentials $\X$ and $\A$, and after that, using Eqs. (\ref{n3}),
(\ref{n4}) let us translate the result to the $\Z$-language. After some
non-trivial algebraic work one obtains that
\be\label{n7}
\M=\D_1^T\M_1\D_1+\D_1^T\M_2\D_2+\D_2^T\M_2^T\D_1+\D_2^T\M_3\D_2-\L,
\ee
where
\be\label{n8}
\M_1=\H^{-1},\quad \M_2=\H^{-1}\Z,\quad \M_3=\Z^T\H^{-1}\Z,
\ee
and
\be\label{n9}
\H=\Sigma-\Z\Xi\Z^T.
\ee
In Eq. (\ref{n7}) the constant matrices $\D_1$ and $\D_2$ read:
\be\label{n10}
\D_1=\left ( \Sigma\,\, 1\,\, 0\right ), \quad
\D_2=\left(
\ba{ccc}
1&-\Sigma&0 \cr
0&0&\sqrt 2
\ea
\right).
\ee
Note, that the block components of $\D_1$ are the $(d+1)\times (d+1)$,
$(d+1)\times (d+1)$ and $(d+1)\times n$ matrices. The first row blocks of $\D_2$
have the same dimensionalities; for the second row one has $n\times (d+1)$,
$n\times (d+1)$ and $n\times n$ blocks. Now let us calculate the vector matrix
$\vec\Omega$ accordingly Eq. (\ref{h12}). In this calculation it is convenient
to use the following multiplication relations, which can be easily established
for the matrices $\D_1$ and $\D_2$:
\be\label{n11}
\D_1\L\D_1^T=2\Sigma,\quad \D_1\L\D_2^T=0, \quad \D_2\L\D_2^T=-2\Xi.
\ee
The result reads:
\be\label{n12}
\vec\Omega=\D_1^T\vec\Omega_1\D_1-\D_1^T\vec\Omega_2\D_2+
\D_2^T\vec\Omega_2^T\D_1+\D_2^T\vec\Omega_3\D_2,
\ee
where
\be\label{n13}
&&\nabla\times\vec\Omega_1=\vec J,\quad \nabla\times\vec\Omega_2=\H^{-1}\nabla\Z-
\vec J\Z, \nonumber\\
&&\nabla\times\vec\Omega_3=\nabla\Z^T\H^{-1}\Z-\Z^T\H^{-1}\nabla\Z
+\Z^T\vec J\Z,
\ee
and the vector current $\vec J$ reads:
\be\label{n14}
\vec J=\H^{-1}\left (\Z\Xi\nabla\Z^T-\nabla\Z\Xi\Z^T\right )\H^{-1}.
\ee
Eqs. (\ref{n7})-(\ref{n10}) and (\ref{n12})-(\ref{n14}) give a translation of
the solution expressed in terms of the $\Z$-related quantities
$(\M_a,\vec\Omega_a)$,\, $a=1,2,3$,\, to the $(\M,\vec\Omega)$-form. The inverse
relations can be easily obtained using the operators
\be\label{n15}
\Pi_1=\frac{1}{2}\L\D_1^T\Sigma,\quad \Pi_2=-\frac{1}{2}\L\D_2^T\Xi.
\ee
The $(\M,\vec\Omega)\rightarrow (\M_a,\vec\Omega_a)$ map reads:
\be\label{n16}
&&\M_1=\Pi_1^T\M\Pi_1+\frac{1}{2}\Sigma,\quad \vec\Omega_1=\Pi_1^T
\vec\Omega\Pi_1,
\nonumber\\
&&\M_2=\Pi_1^T\M\Pi_2,\quad \vec\Omega_2=-\Pi_1^T
\vec\Omega\Pi_2,
\nonumber\\
&&\M_3=\Pi_2^T\M\Pi_2-\frac{1}{2}\Xi,\quad \vec\Omega_3=\Pi_2^T
\vec\Omega\Pi_2.
\ee
In proof of Eq. (\ref{n17}) one can use the following projective properties of
$\Pi_1$ and $\Pi_2$:
\be\label{n17}
&&\D_1\Pi_1=1,\quad \D_1\Pi_2=0,\quad \D_2\Pi_1=0,\quad \D_2\Pi_2=1;
\nonumber\\
&&\Pi_1^T\L\Pi_1=\frac{1}{2}\Sigma,\quad \Pi_1^T\L\Pi_2=0,\quad
\Pi_2^T\L\Pi_2=-\frac{1}{2}\Xi.
\ee
In Eq. (\ref{n16}) the scalar and vector matrices are combined in three doublet
`generations'. The constituents of any doublet have the equivalent matrix
dimensionalities and, moreover, the same natural block structure. This block
structure is induced by Eqs. (\ref{h6}), (\ref{n1}), (\ref{n3}),
(\ref{n4}), (\ref{n8}), (\ref{n9}), (\ref{n13}) and (\ref{n14}). One obtains that
the constituents of the doublets $(\M_1,\vec\Omega_1)$, $(\M_2,\vec\Omega_2)$ and
$(\M_3,\vec\Omega_3)$ are the $2\times 2$, $2\times 3$ and $3\times 3$ block
matrices respectively
with the first row of the following structure: $(1\times 1,\,\,1\times d)$,\,\,
$(1\times 1,\,\,1\times d,\,\,1\times n)$ and
$(1\times 1,\,\,1\times d,\,\,1\times n)$ (the structure of the remaining rows is
completely defined by the given information). This means,
for example, that the `$13$' block component of $\M_2$ (we denote it as
$M_{2,13}$) is the $1\times n$ matrix. Performing the appropriate
segmentation of the matrices $D_1$ and $\D_2$ (then $\D_1$ becomes $2\times 5$,
whereas $\D_2$ is the $3\times 5$ block matrix) and applying Eqs. (\ref{n7}) and
(\ref{n12}), one obtains the explicit relations between the $(\M,\vec\Omega)$
and $(\M_a,\vec\Omega_a)$ block components. The scalar part of the relations
necessary in view of Eq. (\ref{h14}) reads:
\be\label{n19}
&&\M_{11}=\M_{1,11}-2\M_{2,11}+\M_{3,11},\nonumber\\
&&\M_{12}=-\M_{1,12}G_0-\M_{2,12}+(\M_{2,21})^TG_0+\M_{3,12},\nonumber\\
&&\M_{14}=\M_{1,12}-\M_{2,12}G_0+(\M_{2,21})^T-\M_{3,12}G_0,\nonumber\\
&&\M_{15}=\sqrt 2\left (\M_{2,13}+\M_{3,13}\right ),\nonumber\\
&&\M_{22}=G_0\M_{1,22}G_0+G_0\M_{2,22}+(\M_{2,22})^TG_0+\M_{3,22},\nonumber\\
&&\M_{24}=G_0\M_{1,22}-G_0\M_{2,22}G_0+(\M_{2,22})^T-\M_{3,22}G_0-1,\nonumber\\
&&\M_{25}=\sqrt 2\left ( G_0\M_{2,23}+\M_{3,23}\right );
\ee
whereas for the vector components (see Eq. (\ref{h13})) one obtains:
\be\label{n19'}
&&\vec\Omega_{12}=-\vec\Omega_{1,12}G_0+\vec\Omega_{2,12}+(\vec\Omega_{2,21})^TG_0
+\vec\Omega_{3,12},\nonumber\\
&&\vec\Omega_{14}=-\vec\Omega_{1,12}G_0-\vec\Omega_{2,12}G_0+(\vec\Omega_{2,21})^T
-\vec\Omega_{3,12}G_0,\nonumber\\
&&\vec\Omega_{15}=\sqrt 2\left (\vec\Omega_{2,13}+\vec\Omega_{3,13}
\right ),
\ee
where $G_0={\rm diag}(-1;1,...,1)$ is the $d\times d$ matrix which gives the
trivial value of metric components corresponding to the compactified dimensions.
Eqs. (\ref{h13})-(\ref{h16}) and (\ref{n19}), (\ref{n19'}) allow one to transform
the result obtained in the $\Z$-terms to the form of the physical field components.

The last problem in our program is the $\Z$-based description of the hidden
symmetries. First of all, from Eq. (\ref{n5}) it follows that the transformation
\be\label{n20}
\Z\rightarrow C_1\Z C_2
\ee
is a symmetry if
\be\label{n21}
C_1^T\Sigma C_1=\Sigma,\quad C_2^T\Xi C_2=\Xi,
\ee
i.e., if $C_1\in O(2,d-1)$ and $C_2\in O(2,d-1+n)$. It is easy to show that the
realization of this symmetry transformation on the set of the $\Z$-related
quantities $(\M_a,\vec\Omega_a)$ reads:
\be\label{n22}
\ba{ccc}
\M_1\rightarrow C_1^{-1\,T}\M_1C_1^{-1},&
\M_2\rightarrow C_1^{-1\,T}\M_2C_2,&\M_3\rightarrow C_2^T\M_3C_2,\cr
\vec\Omega_1\rightarrow C_1^{-1\,T}\vec\Omega_1C_1^{-1},&
\vec\Omega_2\rightarrow C_1^{-1\,T}\vec\Omega_2C_2,&
\vec\Omega_3\rightarrow C_2^T\vec\Omega_3C_2.
\ea
\ee
Also it is possible to establish the action of the symmetry (\ref{n20}) on the
null-curvature matrix $\M$. Namely, using Eq. (\ref{n7}) and the projective
relations (\ref{n17}), one concludes that
\be\label{n22'}
\M\rightarrow\C^T\M\C,
\ee
where $\C=\C_1\C_2$ and
\be\label{n23}
\C_1=1+\Pi_1(C_1^{-1}-1)\D_1,\quad \C_2=1+\Pi_2(C_2-1)\D_2.
\ee
From Eq. (\ref{n17}) it follows that the matrices $\C_1$ and $\C_2$, which
represent the $\Z$-transformations given by $C_1$ and $C_2$ in the $\M$-terms,
commute and also that both these matrices satisfy the restriction for the group
$O(d+1,d+1+n)$:
\be\label{n24}
\C^T\L\C=\L.
\ee
Eq. (\ref{n24})
contains all hidden symmetries for the theory under consideration
\cite{Sen3}. Of course, the general $O(d+1,d+1+n)$ symmetry transformation does
not coincide with the one derived above. In fact we have detected the subgroup
$O(2,d-1)\times O(2,d-1+n)$ of the complete group of symmetry transformations.
There
is some `missing' symmetry from $O(d+1,d+1+n)$ which action on $\M$ and $\Z$
must be established. For our purposes it will be sufficient to construct it
in the infinitesimal form.

To do it, let us denote the generators of $C_1$ and $C_2$ as $\gamma_1$ and
$\gamma_2$, i.e. let us put $C_1=1+\gamma_1$, \, $C_2=1+\gamma_2$ when
$C_1,\,C_2\rightarrow 1$. For the corresponding generators in the
$\M$-representation from Eq. (\ref{n23}) one has:
\be\label{n25}
\Gamma_1=-\Pi_1\gamma_1\D_1,\quad \Gamma_2=\Pi_2\gamma_2\D_2.
\ee
These generators satisfy the algebra relation which follows from Eq.
(\ref{n24}):
\be\label{n26}
\Gamma^T=-\L\Gamma\L.
\ee
It is easy to prove that the general solution of Eq. (\ref{n26}) can be written
in the form of
\be\label{n27}
\Gamma=\Gamma_1+\Gamma_2+\Gamma_3,
\ee
where
\be\label{n28}
\Gamma_3=2\left (\Pi_2\Xi\gamma_3^T\Pi_1^T-
\Pi_1\gamma_3\Xi\Pi_2^T\right )\L.
\ee
Here $\gamma_3$ is the arbitrary constant parameter, whereas
$\gamma_1^T=-\Sigma\gamma_1\Sigma,\,\,\gamma_2^T=-\Xi\gamma_2\Xi$ in view of Eq.
(\ref{n21}). The generators $\Gamma_a$ define a transformations of the
null-curvature matrix $\M$ of the form (see Eq. (\ref{n23})):
\be\label{n28'}
\delta_a\M=\Gamma_a^T\M+\M\Gamma_a.
\ee
Then, using the $\M_a\leftrightarrow \M$ correspondence (\ref{n16}), it is easy
to prove that the infinitesimal $\Z$-transformations read:
\be\label{n29}
&&\delta_1\Z=\gamma_1\Z,\quad \delta_2\Z=\Z\gamma_2,\nonumber\\
&&\delta_3\Z=\gamma_3-\Z\Xi\gamma_3^T\Sigma\Z.
\ee
Thus, the `missing' part of hidden symmetries moves the trivial $\Z$-value,
i.e., it does not preserve the trivial spatial asymptotics of the fields. From
this
it follows that Eq. (\ref{n20}) gives the general three-dimensional charging
symmetry subgroup. The `missing' transformations must be removed  from the
procedure of the symmetry generation of the asymptotically flat solutions.

Eqs. (\ref{n25}) and (\ref{n28}) show how the operators $\Pi_1$ and $\Pi_2$
realize a relation between the symmetry algebra realizations in $\Z$ and $\M$
representations. Let $T(\gamma_a)$ denotes the infinitesimal transformation in
the arbitrary representation which corresponds to the generator $\gamma_a$. Then,
as it is easy to check, the commutation relations read:
\be\label{n30}
&&[T(\gamma_1^{'},T(\gamma_1^{''})]=T([\gamma_1^{'},\gamma_1^{''}]),\quad
[T(\gamma_2^{'},T(\gamma_2^{''})]=T([\gamma_2^{'},\gamma_2^{''}]),\quad
[T(\gamma_1,T(\gamma_2)]=0;\nonumber\\
&&[T(\gamma_1,T(\gamma_3)]=T(\gamma_3^{'}),\,\,
[T(\gamma_2,T(\gamma_3)]=T(\gamma_3^{''}),\,\,
{\rm where}\,\,
\gamma_3^{'}=-\gamma_1\gamma_3,\,\,\gamma_3^{''}=-\gamma_3\gamma_2;
\nonumber\\
&&[T(\gamma_3^{'},T(\gamma_3^{''})]=T(\gamma_1)+T(\gamma_2),
\,\,{\rm where}\,\,
\gamma_1=(\gamma_3^{''}\Xi\gamma_3^{'\,T}-\gamma_3^{'}\Xi\gamma_3^{''\,T})\Sigma,
\nonumber\\
&&\gamma_2=\Xi(\gamma_3^{'\,T}\Sigma\gamma_3^{''}-
\gamma_3^{''\,T}\Sigma\gamma_3^{'}).
\ee
These relations will play an important role in the study of the
infinite-dimensional symmetry group of the heterotic string theory; this group
arises after the following reduction to two spatial dimensions. In this case
the heterotic string theory becomes the completely integrable two-dimensional
symmetric space model coupled to gravity \cite{Sch}.

\section{Extremal solutions}
\setcounter{equation}{0}

The new approach developed in the previous section gives all the necessary tools
for generation of asymptotically flat solutions of the heterotic string theory
(\ref{h1}) compactified to three dimensions on a torus. This generation procedure
includes the following steps:

\noindent 1).\, One must take some special asymptotically flat solution (or some
consistent heterotic string theory subsystem) and to represent it in terms of the
potential $\Z$ and three-metric $h_{\mu\nu}$.

\noindent 2).\, After that one must apply Eq. (\ref{n20}) with the matrices
$C_1$ and $C_2$ which are the general solutions of the charging symmetry group
relations (\ref{n21}).

\noindent 3).\, One must calculate the charging symmetry transformed values of
the doublets $(\M_a,\vec\Omega_a)$ and after that one must obtain the full set
of the multidimensional field components accordingly the scheme developed in the
two previous sections.

Note, that this simple program realizes the most general technique for generation
of the asymptotically flat solutions in the theory under consideration; this
program leads to generation of the charging symmetry complete classes of
solutions. However, also the $\Z$-formalism can be effectively used for the
straightforward construction of the charging symmetry invariant and asymptotically
flat solution families. In this paper we illustrate this statement by exploring of
one remarkable property of the three-dimensional heterotic string theory. Namely,
there is a close formal analogy between this theory and the classical stationary
Einstein-Maxwell system. Let us clarify this question and give the corresponding
example of application of the developed formalism. So, let us consider the
special subsystem with $d=1$ and $n=2$. Let us separate the $2\times 4$ matrix
potential $\Z$ to the two $2\times 2$ blocks $\Z_{\alpha}$ \, ($\alpha=1,2$,
see Eq. (\ref{n4})),
and define the subsystem under consideration by taking of the anzats
\be\label{e2}
\Z_{\alpha}=
\left(
\ba{cc}
z_{\alpha}^{'}&z_{\alpha}^{''}\cr
-z_{\alpha}^{''}&z_{\alpha}^{'}
\ea
\right).
\ee
It is easy to prove that the motion equations (\ref{n6}) reduce to the system
\be\label{e3}
&&\nabla^2z+2\frac{\nabla z\sigma_3z^+}{1-z\sigma_3z^+}\nabla z=0,
\nonumber\\
&&R_{3\,\mu\nu}=2\frac{z_{,\mu}(\sigma_3-z^+z)^{-1}z^+_{,\nu}}
{1-z\sigma_3z^+},
\ee
where $\sigma_3$ is one of the Pauli matrices, $z=(z_1\,\,z_2)$ and $z_{\alpha}=
z_{\alpha}^{'}+iz_{\alpha}^{''}$. Let us construct some special solution class for
this system. This solution class arises in framework of the anzats
\be\label{e4}
z=\lambda\,q,
\ee
where $\lambda$ is the dynamical complex function and $q$ is the complex constant
$1\times 2$ row. We state that the choice of $\lambda$, $q$ and $h_{\mu\nu}$ such
that
\be\label{e5}
\nabla^2 z=0,\quad q\sigma_3q^+=0,\quad h_{\mu\nu}=\delta_{\mu\nu}
\ee
gives the solution of the equations (\ref{e3}). In fact it is the well-known
Israel-Wilson-Perjes class of solutions \cite{IWP}, and our subsystem is the
conventional stationary Einstein-Maxwell theory. To prove this fact, let us
introduce the potentials
\be\label{e6}
E=\frac{1-z_1}{1+z_1},\quad F=\frac{\sqrt 2z_2}{1+z_1},
\ee
and take the row $q$ in the form of
\be\label{e7}
q=(1\,e^{i\delta}),
\ee
where $\delta$ is real without loss of any generality ($\lambda$ is understood
as the arbitrary complex harmonic function). Then for this solution one has:
\be\label{e8}
E=\frac{1-\lambda}{1+\lambda},\quad F=\frac{e^{i\delta}}{\sqrt 2}(1-E),
\ee
i.e. the formulas defining the general stationary extremal solution class of
the Einstein-Maxwell theory. Also the three-dimensional matter Lagrangian takes
the following form in terms of the new potentials $E$ and $F$:
\be\label{e9}
L_3=L_{EM}=\frac{1}{2f^2}\left | \nabla E-\bar F\nabla F\right |^2-
\frac{1}{f}\left | \nabla F\right |^2,
\ee
where $f=\frac{1}{2}(E+\bar E-|F|^2)$. Thus, $E$ and $F$ are actually the
conventional complex Ernst potentials \cite{ErnstEM}.

In \cite{HSTI} one can find information about the charging-symmetry invariant
generation procedure for the heterotic string theory with the arbitrary values
of $d$
and $n$ starting from this effective Einstein-Maxwell system. Below we  use the
above presented material as the pattern for the straightforward construction of
the general extremal solution class of the Israel-Wilson-Perjes type in the
heterotic string theory. Actually, let us use the explicit similarity between
the systems (\ref{n6}) and (\ref{e3}) and consider the heterotic string theory
anzats
\be\label{e10}
\Z=\Lambda\Q,
\ee
where $\Lambda$ is the dynamical $(d+1)\times \K$ real matrix and $\Q$ is the real
$\K\times (d+1+n)$ matrix constant. It is easy to prove, that the relations
\be\label{e11}
\nabla^2\Z=0,\quad\Q\Xi\Q^T=0,\quad h_{\mu\nu}=\delta_{\mu\nu}
\ee
lead to solution of the motion equations (\ref{n6}). This solution is charging
symmetry complete: an application of the transformation (\ref{n20}) is equivalent
to the re-parameterization $\Lambda\rightarrow C_1\Lambda$, \,
$\Q\rightarrow\Q C_2$. This re-parameterization is non-important because there is
not any algebraical restriction on $\Lambda$ and the restriction on $\Q$ is
invariant in respect to this re-parameterization (of course, we take the general
matrix $\Q$ satisfying Eq. (\ref{e11})).

Let us now obtain the heterotic string theory analogy of Eq. (\ref{e7}). In
particular, we would like to clarify the question about the concrete possible
values of the parameter $\K$. First of all, from Eq. (\ref{e10}) it follows that
the $\Q$-rows can be taken algebraically independent. Actually, if, for example,
the $\K$-th row is a linear combination of the other ones, i.e. if $\Q_{\K}=
\sum_{l=1}^{\K-1}\beta_l\Q_l$ \, ($\beta_l$ are the coefficients), then the
removing of $\Q_{\K}$ from $\Q$ together with the replacement of the $l$-th column
$\Lambda_l$ of $\Lambda$ to the column $\Lambda_l+\beta_l\Lambda_{\K}$ effectively
transforms Eq. (\ref{e10}) to the same one but with the shift $\K\rightarrow\K-1$.
Thus, we can actually take $\Q$ with the independent rows without loss of
generality.
Then, from this it follows that $\K<d+1+n$ (because the numbers of the algebraically
independent rows and columns coincide for any matrix and in the case of
$\K=d+1+n$ one has from Eq. (\ref{e11}) that ${\rm det}\,\Q=0$, i.e. the
contradiction with the proposed matrix $\Q$ row independence). As result it exists the
parameterization of $\Q$ of the following form:
\be\label{e12}
\Q=(\Q_1\,\Q_2),
\ee
where $\Q_1$ and $\Q_2$ are the $\K\times\K$ and $\K\times (d+1+n-\K)$ matrices
respectively.
The matrix $\Q_1$ must be non-degenerated ($\Q$ has the independent rows), so one
can represent $\Q$ as $\Q=\Q_1(1\,\,\N)$, where $\N=\Q_1^{-1}\Q_2$. It is easy to
see that the matrix $\Q_1$ can be absorbed by the re-parameterization
$\Lambda\Q_1\rightarrow\Lambda$ without loss of generality, so the pair
$(\Lambda,\,\N)$ defines our solution completely. Then, it is not difficult to
establish that the algebraical $\Q$-restriction in Eq. (\ref{e11}), being written
in terms of the established new form of the matrix $\Q$,
\be\label{e13'}
\Q=(1\,\,\,\N),
\ee
is compatible only if $\K=1,2$. In the case of $\K=1$ the matrix
$\Lambda=\Lambda^{(1)}$ is the
$(d+1)\times 1$ column, whereas $\N=\N^{(1)}$ is the $1\times (d+n)$ row. This row
satisfies the following algebraical restriction:
\be\label{e13''}
\bar\N^{(1)}\bar\N^{(1)\,T}=1+(\tilde\N^{(1)})^2,
\ee
where the parameterization $\N^{(1)}=(\tilde\N^{(1)}\,\,\bar\N^{(1)})$ is
performed (here $\tilde\N^{(1)}$ is the number and $\bar\N^{(1)}$ is the
$1\times (d-1+n)$ row).
In the case of $\K=2$ \,\, $\Lambda=\Lambda^{(2)}$ is the $(d+1)\times 2$ matrix,
whereas $\N=\N^{(2)}$ consists of two $1\times (d-1+n)$ mutually orthogonal rows
$\N_1^{(2)},\,\N_2^{(2)}$ of the unit norm:
\be\label{e13}
\N^{(2)}=
\left(
\ba{c}
\N_1^{(2)}\cr
\N_2^{(2)}
\ea
\right),
\quad
\N_1^{(2)}\N_1^{(2)\,T}=\N_2^{(2)}\N_2^{(2)\,T}=1,\,\,\N_1^{(2)}\N_2^{(2)\,T}=0.
\ee
The case of $\K=1$ exists for the values of $(d,n)$ such that $d+n\geq 2$
(it follows from Eq. (\ref{e13''})), whereas the case of $\K=2$ is consistent if
$d+n\geq 3$. Thus, the first case is the single possible one for the theories with
$d=n=1$ \, (the four-dimensional Einstein-Maxwell dilaton-axion theory
\cite{EMDA}) and with $d=2,\,n=0$ \, (the five-dimensional dilaton-Kalb-Ramond
gravity see or its on-shell equivalent analogy (\cite{Kallosh})). The special
classes of extremal solutions of these theories one can find in \cite{OKMY}. In
fact only in these two special cases the extremal solution with $\K=1$ exists as
the original solution: we state that for all other situations, i.e. when
the condition of the solution with $\K=2$ is satisfied, the $\K=1$ solution
branch is some special case of the branch with $\K=2$. Actually, the submersion of
the former branch into the latter one has the extremely simple form:
\be\label{e14}
\Lambda_1^{(2)}=\Lambda^{(1)},\,\,\,\,\Lambda_2^{(2)}=\tilde\N^{(1)}\Lambda^{(1)},
\,\,\,\,\N^{(2)}_1+\tilde\N^{(1)}\N_2^{(2)}=\bar\N^{(1)}.
\ee
These relations mean the following: $\Lambda^{(1)}$, \, $\tilde\N^{(1)}$ and $\bar
\N^{(1)}$ are the arbitrary parameters (they parameterize the branch with
$\K=1$), whereas $\Lambda^{(2)}_1$, $\Lambda^{(2)}_2$, $\N^{(2)}_1$ and
$\N^{(2)}_2$ must be taken to satisfy Eq. (\ref{e14}). Note, that the last
relation in (\ref{e14}) can always be realized; for the proof one can use some
simple geometrical reasons. Namely, $\bar\N^{(1)}$ is the vector in the at least
two-dimensional Euclidean space $(d-1+n\geq 2)$ with the norm
$\sqrt{1+(\tilde\N^{(1)})^2}$. It is clear that one always can take the
two-dimensional plane in this space and two unit mutually orthogonal vectors
$\N^{(2)}_1$
and $\N^{(2)}_2$ in it in such a way that the vector $\bar\N^{(1)}$ has the unit
projection on $\N^{(2)}_1$ and the projection on $\N^{(2)}_2$ equal to
$\tilde\N^{(1)}$. This choice exactly corresponds to the third relation in Eq.
(\ref{e14}). Both the critical heterotic and bosonic string theories are the
consistent representatives for the our extremal solution class with $\K=2$.

Let us now briefly discuss the solution decoding procedure in this case, i.e.
the translation of the found extremal solution from the $\Z$-form to the
language of the multidimensional string
theory fields. In this procedure the single qualitative step is related to
calculation of the matrices $\M_a$ and $\vec\Omega_a$; all the remaining work is
principally simple but technically tedious algebra. The calculation gives:
\be\label{e15}
\ba{ccc}
\M_1=\Sigma,&\M_2=\Sigma\Lambda\Q,&\M_3=\Q^T\Lambda^T\Sigma\Lambda\Q,\cr
\vec\Omega_1=0,&\vec\Omega_2=\Sigma\vec\Theta_1\Q,&
\vec\Omega_3=\Q^T\vec\Theta_2\Q,
\ea
\ee
where the vector fields $\vec\Theta_1$ and $\vec\Theta_2$ are defined by the
relations
\be\label{e16}
\nabla\times\vec\Theta_1=\nabla\Lambda,\quad
\nabla\times\vec\Theta_2=\nabla\Lambda^T\Sigma\Lambda-
\Lambda^T\Sigma\nabla\Lambda.
\ee
For the solutions with $\K=1$ the term $\vec\Theta_2$ vanishes; for the solutions
with $\K=2$ this term is the $2\times 2$ antisymmetric matrix, its
$`12'$-component is defined by the relation
\be\label{e17}
\nabla\times\vec\Theta_{2,12}=\nabla\Lambda_1^{(2)\,T}\Sigma\Lambda_2^{(2)}-
\nabla\Lambda_2^{(2)\,T}\Sigma\Lambda_1^{(2)}.
\ee
We will not continue the calculation of the heterotic string theory field
components in this paper. Let us only note, that the constructed solution class is
asymptotically flat if the harmonic matrix function $\Lambda$ vanishes at the
spatial infinity. In the case when $\Lambda$ has the Coulomb asymptotic behavior
the vector matrix $\vec\Omega_1$ generates the Dirac string peculiarities,
whereas
$\vec\Omega_2$ leads to the dipole moments of the fields. However, this simple
picture is not complete, there are some nuances and not principal moments.
In the
forthcoming publication we hope to continue the analysis of the above established
Israel-Wilson-Perjes type solution of heterotic string theory and also to give
some new analytic material related to its supersymmetric properties.

\section{Conclusion}
\setcounter{equation}{0}

In this paper we have developed a new formalism for the low-energy heterotic
string theory compactified to three dimensions on a torus. This formalism is
extremely compact and based on the use of the single matrix potential $\Z$. The
formalism includes three pairs of the $\Z$-related quantities $(\M_a,
\vec\Omega_a)$; they play an important role in the translation of the
$\Z$-expressed solution to the language of the heterotic string theory field
components. Using our formalism, one can generate asymptotically flat solutions of
the theory which possess the charging symmetry completeness property. The positive
feature of the new approach is related to the fact that the charging symmetry
subgroup of transformations of the theory acts as linear homogeneous map on the
matrix potential $\Z$. Then, the straightforward construction of new solutions
seems at least really promising in this approach. We have illustrated this
statement by the straightforward construction of the general extremal solution
of the Israel-Wilson-Perjes type for the  heterotic string theory with arbitrary
numbers
$d$ and $n$ of the toroidally compactified dimensions and original
multidimensional Abelian gauge fields. For the case of $d=n=1$ our class is
defined by $2$ arbitrary harmonic functions and $1$ constant parameter; for $d=2,
n=0$ we have $3$ harmonic and $1$ constant parametric degrees of freedom. For the
remaining heterotic string theory cases one has $2(d+1)$ basic arbitrary harmonic
functions and $2(d+n)-5$ independent constant parameters. We state, that for the
special subset of $d=1$ and $n\geq 2$ theories our solution exactly coincides with the
solution obtained in \cite{KalIWP}, if one removes all the spatial field
asymptotics from the latter solution. Of course, we can also generate them in our
solution using the shift transformation (with the generator $\gamma_3$ in
$\Z$-representation).

Concerning the nearest perspectives of the activity based on the use of the new
formalism, we hope to apply it in combination with the inverse scattering
transform method for construction of the general two-dimensional soliton solution
of the theory. The three-dimensional generation using the total subgroup of the
charging symmetry transformations accordingly the plan formulated in the previous
section is also in our plans. Also it is necessary to perform the supersymmetric
analysis of the Israel-Wilson-Perjes type solution constructed in this article. It
seems, that there is a compact supersymmetric generalization of the $\Z$-based
formalism, but the corresponding work is now in its early beginning.


\section*{Acknowledgments}
This work was supported by RFBR grant ${\rm N^{0}}
\,\, 00\,02\,17135$.


\end{document}